\documentclass{elsart5p}

\usepackage{graphics}
\usepackage{graphicx}
\usepackage{amssymb}
\begin{document}

\begin{frontmatter}


\title{\textbf{Magnetic properties and magnetocaloric effect in a Laves phase compound ~$\mbox{Er}_2\mbox{Rh}_{3}\mbox{Si}$}}

\author{Baidyanath Sahu}, and 
\author{Andr\'{e} M. Strydom}

\address{Highly Correlated Matter Research Group, Department of Physics, University of Johannesburg, PO Box 524, Auckland Park 2006, South Africa}

\corauth[abd]{Corresponding author. baidyanathsahu@email.com}

\begin{abstract}
The magnetic properties and magnetocaloric effect of an intermetallic compound $\mathrm{Er_2Rh_3Si}$ was investigated by performing temperature dependence of dc$\textendash$ and ac$\textendash$magnetic susceptibility and isothermal magnetizations. Polycrystalline $\mathrm{Er_2Rh_3Si}$ compound with $\mathrm{Mg_2Ni_3Si}$-type of rhombohedral Laves phases (space group $R\overline{3}m$, hR18) structure was synthesized by arc-melting method. The ac$\textendash$ and dc$\textendash$magnetic susceptibility results revealed that the compound exhibited ferromagnetic order with $T_C$ $\approx$ 18 K. The obtained isothermal magnetic entropy changes ($\Delta S_m$), and relative power cooling for a change of magnetic field 0$\textendash$9 T are 10.1~$\mathrm{J.~kg^{-1}.~K^{-1}}$, and 255 J. kg$^{-1}$ respectively. Arrott plots and also the universal scaling plot by rescaled temperature axis and normalized $\Delta S_m$ collapse onto a one curve for independent of magnetic fields, confirming a second order ferromagnetic to paramagnetic phase transition in $\mathrm{Er_2Rh_3Si}$.

\end{abstract}

\begin{keyword}
Ferromagnet; Magnetic susceptibility ; Universal scaling plot; Magnetocaloric effect 

\end{keyword}

\end{frontmatter}


\section{INTRODUCTION}

Study of magnetocaloric effect (MCE) of a magnetic material is an important for the application in solid state magnetic refrigeration technology. Solid state magnetic refrigeration has advantages over the commercial gas compression/expansion as it is environmental friendly. 

MCE is an intrinsic magneto-thermodynamic phenomena in a magnetic material, which describes the changing of temperature with applicant of external magnetic field in an adiabatic environment. MCE materials can be classified from the characteristics value of the isothermal magnetic entropy change ($\Delta S_m$), adiabatic temperature change ($\Delta T_{ad}$), and relative power cooling (RCP) by changing of magnetic field \cite{Book,review,Buschow,RTX,R2T2X}. Research on MCE of rare earth based ternary intermetallics magnetic compounds are devoted low temperature application in magnetic refrigerator due to the relatively low transition temperatures \cite{RTX,Tm2NiSi3}. 

Rhombohedral Laves phases for binary ($\mathrm{RT_2}$; R = rare-earths elements and T = transition metals) have attracted interest due to their fantastic magnetic properties and magnetocaloric effect \cite{RCo2, RNi2, RAl2, RFe2}. Our previous studies reveals that ternary Laves phases ($\mathrm{R_2T_3X}$; X = S and P block elements) intermetallic compounds can posses good MCE \cite{Gd2Rh3Ge,Tb2Rh3Ge}. It is also reported that Eribium (Er) based ternary intermetallic compounds are attractive for the large/giant MCE properties \cite{ErRuSi,ErRu2Si2,Er2Cu2In}.  In this paper, a systematic magnetic study is presented in details for the understanding of magnetic properties in the ternary Laves phases compound, $\mathrm{Er_2Rh_3Si}$. The MCE of $\mathrm{Er_2Rh_3Si}$ has been investigated using magnetic isotherms. A second order ferromagnetic to paramagnetic phase transition has been observed in $\mathrm{Er_2Rh_3Si}$.

\section{EXPERIMENTAL DETAILS}
Polycrystalline sample of $\mathrm{Er_2Rh_3Si}$ was synthesized by arc$\textendash$melting technique on a water cooled copper hearth under high purity argon atmosphere. High purities constituent elements (purities are 99.99 wt \%) were used for this sample. The ingot was flipped and remelted several times to ensure homogeneous mixing of the constituents. The weight loss after final melting was less than 0.5~ wt.\%. The ingot was wrapped with tantalum foil, encapsulated in an evacuated quartz tube and annealed at 1273 K for one week and then quenched in cold water. The room temperature powder X-ray diffraction (XRD) patterns was collected using Cu$\textendash$K$_\alpha$ radiation of Rigaku X-ray diffractometer. The Powder XRD pattern was analyzed by Rietveld refinement method using the ‘‘FULLPROF” software \cite{rietveld,fullprof}. Both dc$\textendash$ and ac$\textendash$ magnetic measurements were performed on a commercial Dynacool physical property measurement system (PPMS), attached with a vibrating sample magnetometer (VSM) option (made by Quantum Design, USA). Temperature dependence of dc$\textendash$magnetization ($M(T)$) was measured under zero field cooled (ZFC) and field cooled (FC) protocol \cite{Gd2Rh3Ge}.

\section{RESULTS AND DISCUSSIONS}
\subsection{X-ray diffraction}

\begin{figure}[!t]
	\includegraphics[scale=0.35]{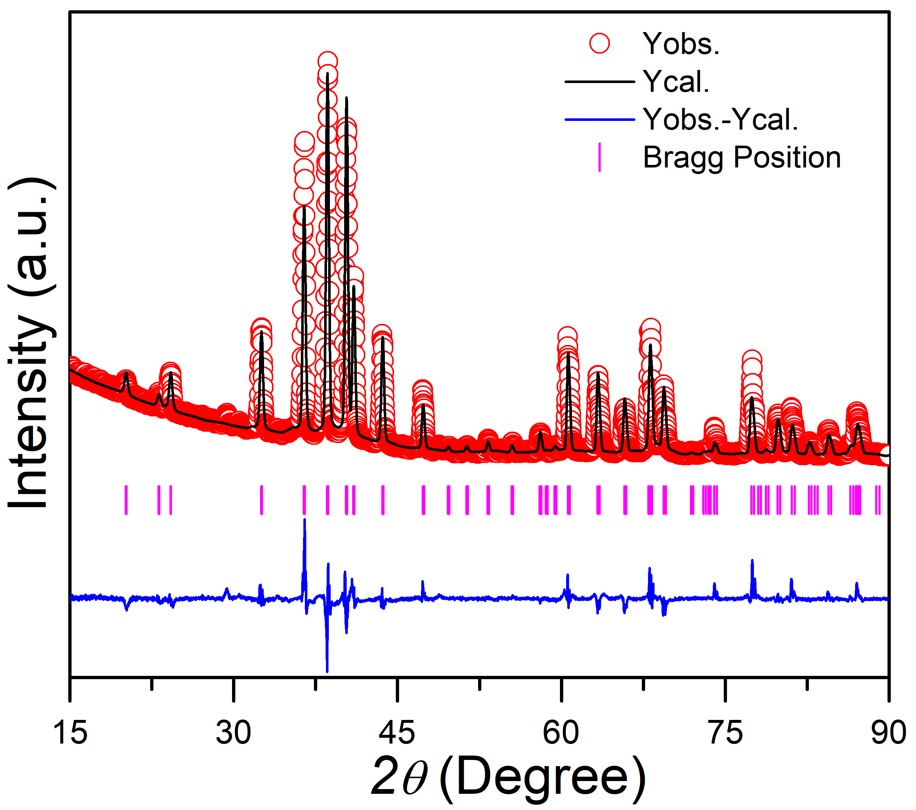}
	\caption{The experimentally observed powder XRD pattern of $\mathrm{Er_2Rh_3Si}$ along with the Rietveld refinement fit using space group $R\overline{3}m$, hR18.}
	\label{XRD}
\end{figure}

Fig.~\ref{XRD} shows the experimental and refined powder XRD patterns of $\mathrm{Er_2Rh_3Si}$ compound at room temperature. The Rietveld analysis was  done using a rhombohedral phase with space group $R\overline{3}m$. The Rietveld refinement analysis revealed that $\mathrm{Er_2Rh_3Si}$ crystallizes in $\mathrm{Mg_2Ni_3Si}$$\textendash$type of rhombohedral Laves phases. The  obtained lattice parameters of a unit cell are a = b = 5.556(2) \AA  and c =  11.829(3) \AA. These obtained lattice parameters are in good agreement with previously report values \cite{Y2Rh3Ge}.

\subsection{Magnetic  properties}

\begin{figure}[!t]
	\includegraphics[scale=0.34]{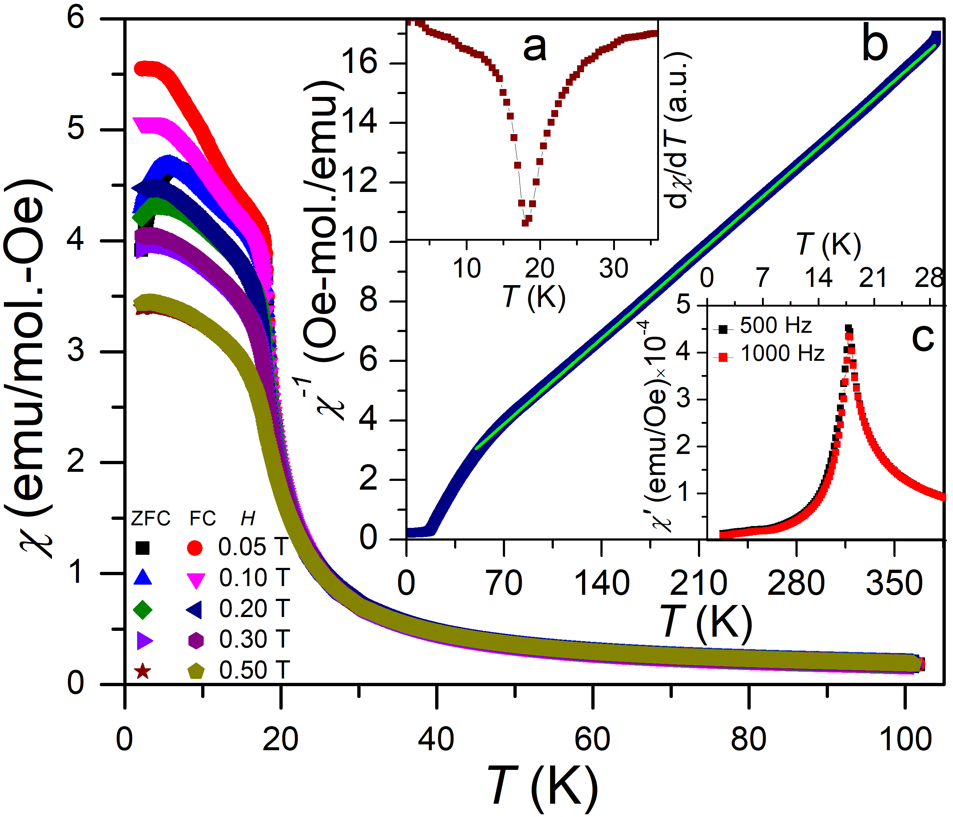}
	\caption{Temperature dependence of ZFC and FC dc$\textendash$magnetic susceptibility ($\chi(T)$) under different external magnetic field. Inset (a): d$\chi(T)$/d$T$ of the FC curve for magnetic field of 0.5 T. Inset (b): Temperature variation of inverse $\chi(T)$ of the FC curve for external field of 0.2 T along with the Curie$\textendash$Weiss fitting line. Inset (c): Temperature variation real part of ac$\textendash$susceptibility for two different frequencies 500 and 1000 Hz, while applied ac$\textendash$driven field 0.3 mT and zero external dc$\textendash$magnetic field.}
	\label{MT}
\end{figure}

Fig.~\ref{MT} shows the temperature dependence of ZFC and FC dc$\textendash$magnetic susceptibility ($\chi(T)$ = $M(T)/H$) for different applied magnetic fields. The ZFC and FC of $\chi(T)$ for all fields show a typical paramagnetic to ferromagnetic (PM-FM) transition. The Curie temperatures ($T_C$)~$\approx$~18 K was derived from the peak value of d$\chi(T)$/d$T$~$vs.$~$T$ of FC curves (shown in inset(a) of Fig.~\ref{MT}). Below $T_C$, the ZFC and FC show a kink at 4 K, which is probably due to the spin reorientation at low temperature in $\mathrm{Er_2Rh_3Si}$. The external field dependent thermomagnetic irreversibility below $T_C$ is observed in ZFC and FC dc$\textendash$magnetic susceptibility curves. This thermomagnetic irreversibility may be as a result of the magnetic anisotropy and the domain wall pinning effect in $\mathrm{Er_2Rh_3Si}$ \cite{Pr2Rh3Ge,book1}.

Inset(b) of Fig.~\ref{MT} shows the inverse dc$\textendash$magnetic susceptibility ($\chi^{-1}(T)$) under an applied field of 0.2 T. The linear behavior of $\chi^{-1}(T)$ deviates below 80 K. This deviation may be due to the influenced of crystalline field of $\mathrm{Er^{3+}}$. The $\chi^{-1}(T)$ of T $\textgreater$ 80 K, was expressed by Curie-Weiss law  $\chi^{-1}(T)$ = (${T-\theta_P}$)/C, where $\mathrm{C = N \mu^{2}_{eff}/3k_{B}}$ is known as Curie constant and $\theta_P$ is the Weiss paramagnetic temperature. The fitting line of Curie$\textendash$Weiss law to the experimental data is shown in inset(b) of Fig.~\ref{MT}. The least$\textendash$squares (LSQ) fit on experimental data yielded a parameter $\theta_P$ = -24.16 K. The observed negative value reveals that antiferromagnetic interaction are associated with ferromagnetic ordering in $\mathrm{Er_2Rh_3Si}$. The calculated $\mathrm{\mu_{eff}}$ = 9.85 $\mathrm{\mu_{B}}$/$\mathrm{Er}$. This obtained $\mathrm{\mu_{eff}}$ is slightly higher than the  the theoretical value for a free trivalent $\mathrm{Er^{3+}}$, $\mathrm{g_{J}[J(J + 1)]^{1/2}}$ = 9.58 $\mathrm{\mu_{B}}$ for J = 15/2, may be the result of crystal field effect (CEF). The obtained $\mathrm{\mu_{eff}}$ value indicate that 4f$\textendash$ shell electrons of $\mathrm{Er^{3+}}$ ions are the predominant magnetic species for $\mathrm{Er_2Rh_3Si}$.

To further confirm the magnetic transition, ac$\textendash$susceptibility was measured. Inset(c) of Fig.~\ref{MT} shows the real part of the $\chi'_{ac}(T)$ data recorded at two different ac$\textendash$frequencies; 500 and 1000 Hz. The $\chi'_{ac}(T)$ exhibits a single peak at 18 K due to ferromagnetic transition. There is no significant difference is observed in $\chi'_{ac}(T)$ for two different frequencies, indicating that there no signature of spin or cluster glass behavior in $\mathrm{Er_2Rh_3Si}$.

\begin{figure}[!t]
	\includegraphics[scale=0.33]{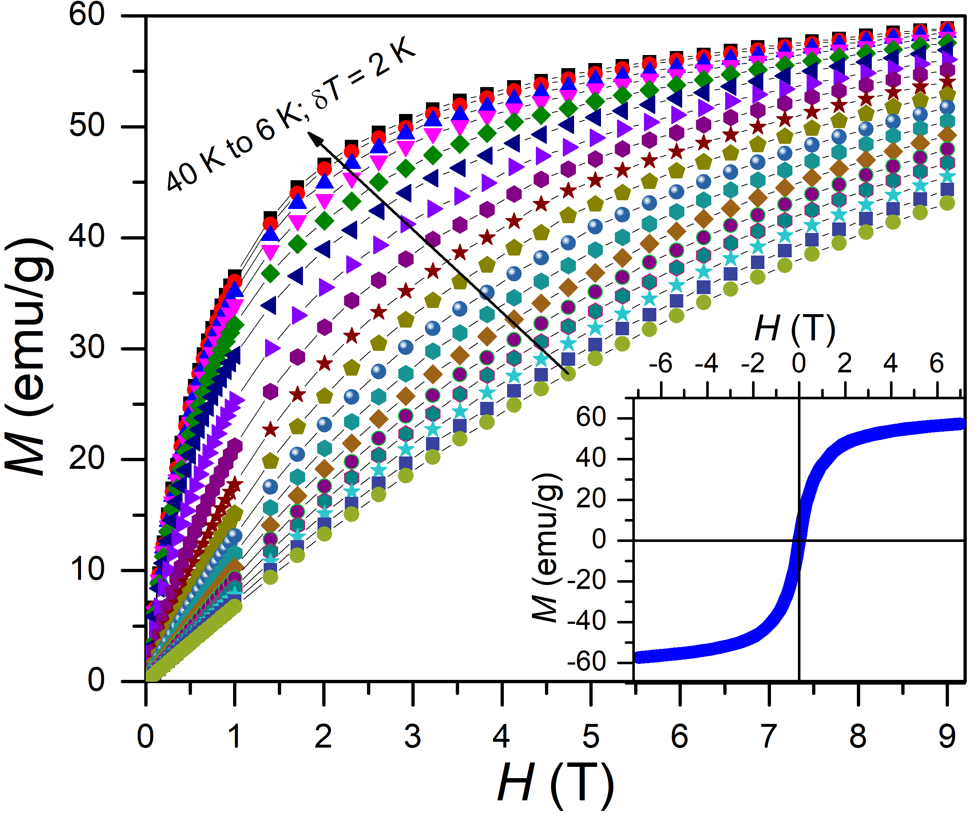}
	\caption{Magnetic field dependence of isothermal magnetization at different temperature from 6 K to 40 K with a step of 2 K. Inset; field dependence magnetization loop at 2 K.}
	\label{MH}
\end{figure}

To investigate the order of magnetic phase transition and $\Delta S_m$, isothermal magnetization ($M(H)$) were measured at different temperature upto a field value of 9 T. The recorded $M(H)$ is shown in Fig.~\ref{MH}. Inset of Fig.~\ref{MH}, shows the magnetization loop at 2 K upto a field value of $\pm$7 T. There no hysteresis is observed in $M(H)$ at 2 K, indicating the $\mathrm{Er_2Rh_3Si}$ is a is a soft ferromagnetic material, which is beneficial for the practical application in magnetic refrigerator \cite{Dy2Cu2In}. One can see that there is no tendency for saturation in the measured range. The spontaneous magnetization was estimated by extrapolating higher field part to zero external field yielding a value of 24.1 (emu/g)/Er = 2.9 $\mu_B$/Er. The magnetization value at 7 T is 28.8 (emu/g)/Er = 3.5 $\mu_B$/Er for 2 K. The obtained magnetization is smaller than the magnetic moment of a free $\mathrm{Er^{3+}}$ ion (g$J$ = 9 $\mu_B$). The reduction of saturation value may result from the magnetic anisotropy due to the CEF effects in $\mathrm{Er_2Rh_3Si}$.

\begin{figure}[!t]
	\includegraphics[scale=0.325]{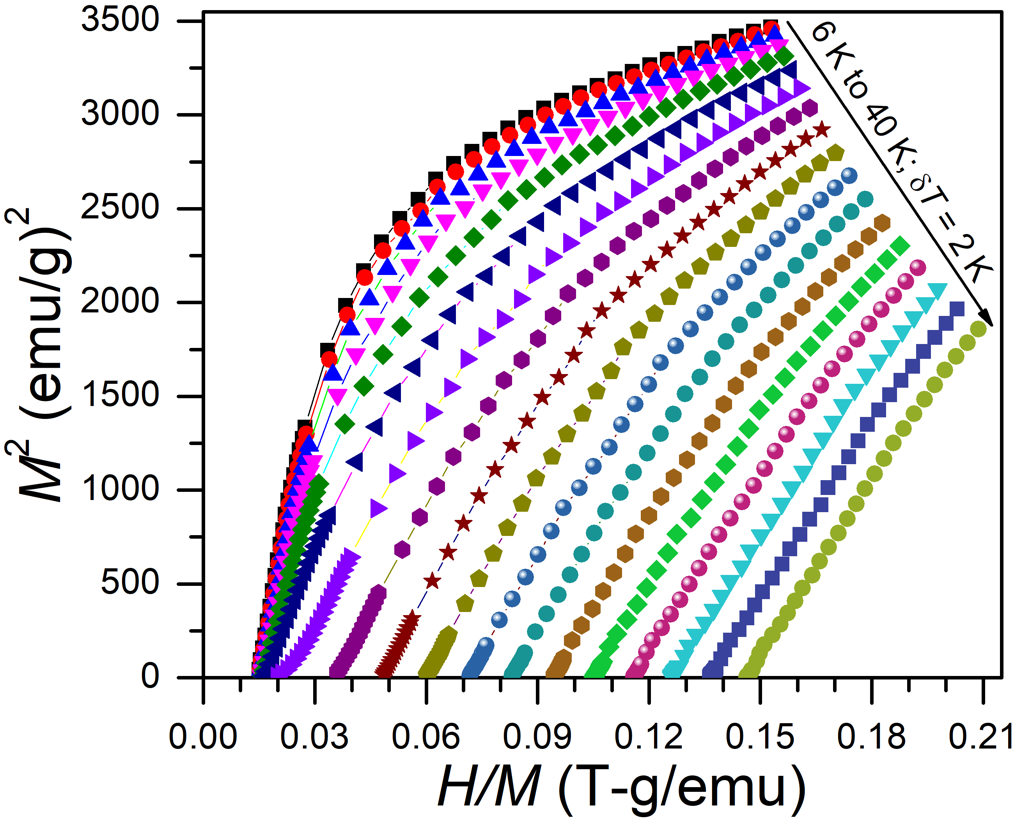}
	\caption{Arrott plots of $M^2$~$vs.$~$H/M$ at different temperatures ranging from 6 K to 40 K with a step of 2 K.}
	\label{AP}
\end{figure}

In order to get more information about nature of order of magnetic phase transition, Arrott plots ($M^2~vs.~H/M$) were made from the isothermal magnetization and is shown in Fig.~\ref{AP}. As per the Banerjee criterion \cite{Banarjee}, the positive slops in $M^2~vs.~H/M$ indicate $\mathrm{Er_2Rh_3Si}$ compound undergoes second order ferromagnetic to paramagnetic phase transition. The order of magnetic phase transitions is also confirmed by employing universal scaling plots of normalized magnetic entropy change as discussed later.

\subsection{Magnetocaloric effect}

\begin{figure}[!t]
	\includegraphics[scale=0.335]{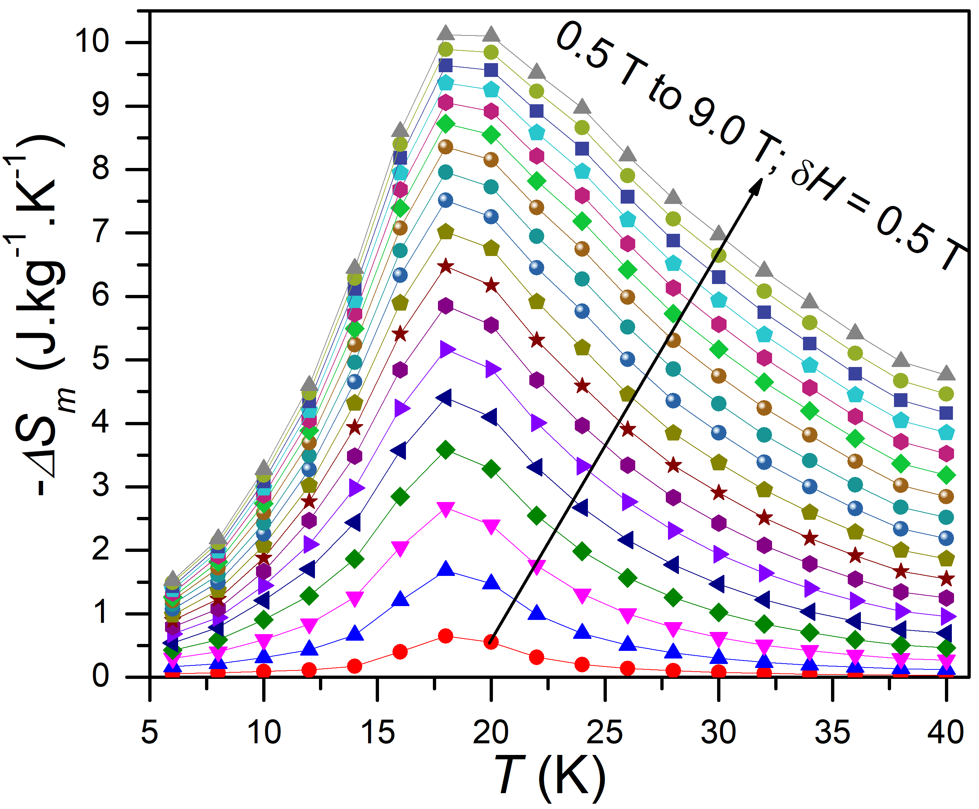}
	\caption{Temperature variation of isothermal magnetic entropy changes ($\Delta S_m$) of $\mathrm{Er_2Rh_3Si}$ for different value of magnetic field.}
	\label{MCE}
\end{figure}

The $\Delta S_m$ of $\mathrm{Er_2Rh_3Si}$ was calculated from the isothermal magnetization curves by using the following Maxwell's magnetic thermodynamic relation \cite{Book,review}
\begin{equation}
\Delta S_m (T, H) = \int_{H_i}^{H_f} \left(\frac{\partial M}{\partial T}\right)_{H'} dH',
\label{DeltaS}
\end{equation}
where $H_i$ is the initial magnetic field (zero in this present study) and $H_f$ is the final field (9 T in this present experiment). Fig.~\ref{MCE} shows the temperature variation $- \Delta S_m$ for different values of applied fields. The obtained $- \Delta S_m$ values are positive for a wide range of temperatures and a peak appears around the $T_C$, indicating paramagnetic to ferromagnetic transition. The maximum value of isothermal magnetic entropy changes $(- \Delta S^{max}_m$) gradually increases with increasing magnetic field. It is found that the $- \Delta S^{max}_m$ values are 3.6 , 7.5, 9.1 and 10.1 $\mathrm{J.kg^{-1}.K^{-1}}$ for the change of magnetic field from 0$\textendash$2, 0$\textendash$5, 0$\textendash$7 and 0$\textendash$9 T respectively.

The relative cooling power (RCP) is also another important parameter to quantify the heat transfer between the hot and cold reservoirs \cite{RCP}. The RCP is defined as the product of the maximum magnetic entropy change $|- \Delta S^{max}_m|$ and the full width at half maximum ($\delta T_{FMHM})$ of  $\Delta S_m$~$vs.$~$T$ curves; RCP~=~$|- \Delta S^{max}_m|$~$\times$~$\delta T_{FMHM}$ \cite{RCP}. The RCP was estimated from $\Delta S_m$~$vs.$~$T$ curves for different magnetic fields. The obtained RCP values gradually increase with the increase of change in magnetic field. The RCP values are 40, 135, 200, and 255 J.kg$^{-1}$ for a change of magnetic field 0$\textendash$2, 0$\textendash$5, 0$\textendash$7, and 0$\textendash$9 T respectively. The magnetic transition temperature, $- \Delta S^{max}_m$ and RCP values at 5 T of $\mathrm{Er_2Rh_3Si}$ and reported Er$\textendash$based ternary intermetallic magnetic compounds are listed in Table~\ref{tableMCE} for the sake of comparison. From these obtained $- \Delta S^{max}_m$ and RCP value, it can be assumed that $\mathrm{Er_2Rh3Si}$ belongs to a class of considerable MCE materials.

\begin{table}[!t]
\caption{\label{tableMCE} The transition temperature ($T_M$), the maximum value of magnetic entropy change ($\Delta S_m^{max}$), relative cooling power, RCP, under the change of magnetic field 0$\textendash$5~T for $\mathrm{Er_2Rh_3Si}$ together with ternary Er$\textendash$based intermetallic compounds. }
\begin{tabular}{ccccc} \\ \hline
\hspace{-0.01 in}Compound	 & 	\hspace{0.17 in} $T_M$	 & 	\hspace{0.17 in}$-\Delta S_m$	 &	 \hspace{0.17 in}RCP	 &		 \hspace{0.17 in}Ref  \\
\hspace{-0.01 in} & \hspace{0.17 in} (K) & \hspace{0.17 in} ($\mathrm{J.kg^{-1}.K^{-1}}$) & \hspace{0.17 in} (J.kg$^{-1}$) & \hspace{0.17 in}   \\ \hline	
\hspace{-0.01 in}$\mathrm{Er_2Rh_3Si}$ & \hspace{0.17 in} 18 & \hspace{0.17 in} 7.5 & \hspace{0.17 in} 135 & \hspace{0.17 in} PW  \\
\hspace{-0.01 in}$\mathrm{Er_2Rh_3Ge}$ & \hspace{0.17 in} 21 & \hspace{0.17 in} 9.2 & \hspace{0.17 in} 225 & \hspace{0.17 in} \cite{Gd2Rh3Ge}  \\
\hspace{-0.01 in}$\mathrm{Er_2Co_2Al}$ & \hspace{0.17 in} 32/21 & \hspace{0.17 in} 5.9 & \hspace{0.17 in} 152 & \hspace{0.17 in} \cite{Er2Co2Al}  \\
\hspace{-0.01 in}$\mathrm{Er_2Ni_2Ga}$ & \hspace{0.17 in} 7.1 & \hspace{0.17 in} 9.62 & \hspace{0.17 in} 196 & \hspace{0.17 in} \cite{Er2Ni2Ga}  \\
\hspace{-0.01 in}$\mathrm{ErRhSi}$ & \hspace{0.17 in} 8.5 & \hspace{0.17 in} 8.7 & \hspace{0.17 in} --- & \hspace{0.17 in} \cite{ErRhSi}  \\
\hspace{-0.01 in}$\mathrm{ErAgAl}$ & \hspace{0.17 in} 14 & \hspace{0.17 in} 10.5 & \hspace{0.17 in} 261 & \hspace{0.17 in} \cite{ErAgAl}  \\
\hspace{-0.01 in}$\mathrm{ErPtGa}$ & \hspace{0.17 in} 5 & \hspace{0.17 in} 8.0 & \hspace{0.17 in} 159 & \hspace{0.17 in} \cite{ErPtGa}  \\
\hspace{-0.01 in}$\mathrm{ErFeAl}$ & \hspace{0.17 in} 55 & \hspace{0.17 in} 6.1 & \hspace{0.17 in} 311 & \hspace{0.17 in} \cite{ErFeAl}  \\
\hline
PW: Present Work	
\end{tabular}	
\end{table}

\begin{figure}[!t]
	\includegraphics[scale=0.315]{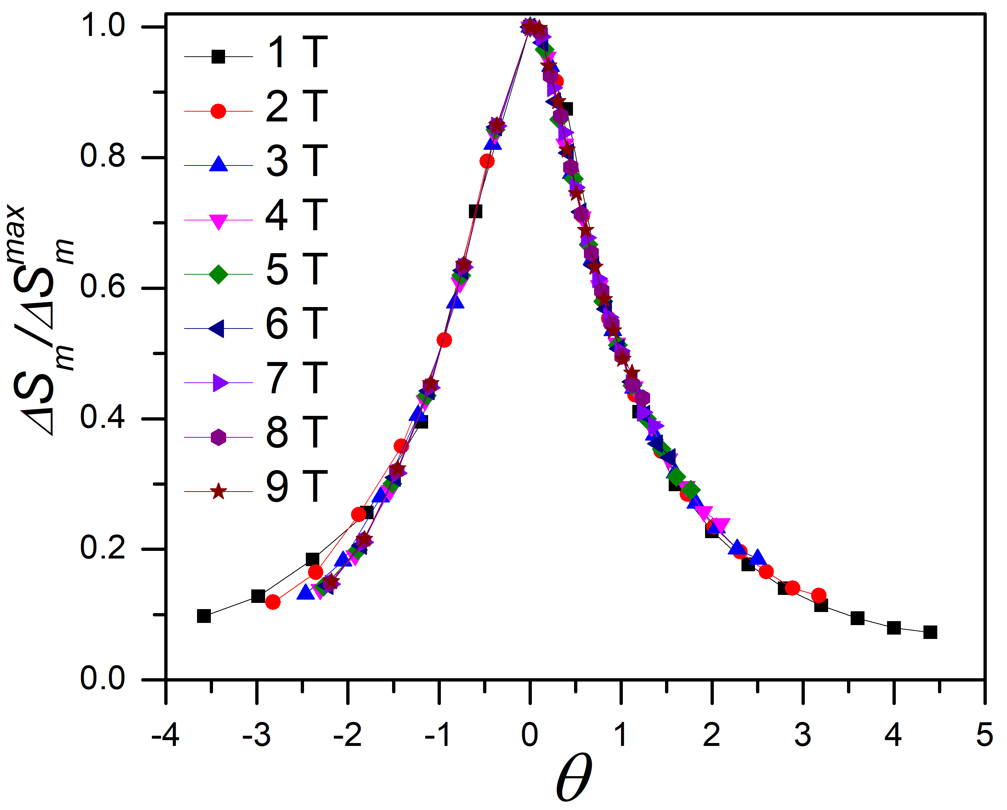}
	\caption{Universal scaling plot by normalized isothermal magnetic entropy change ($\Delta S_m$/$\Delta S^{max}_m$) as a function of the rescaled temperature $\theta$ around $T_C$ for different values of magnetic field of $\mathrm{Er_2Rh_3Si}$.}
	\label{scaling}
\end{figure}

Franco $et$ $al.,$ \cite{franco2006scaling,franco2008universal,franco2010scaling,bonilla2010new,bonilla2010universal} have proposed an universal scaling plot by normalizing temperature dependent isothermal magnetic entropy change for a second order magnetic phase transition material. Universal scaling curves are constructed by the normalized entropy change ($\Delta S_m$/$\Delta S^{max}_m$) against rescaled temperature ($\theta$). The $\theta$ is given by the following expression; 
 \begin{eqnarray} \nonumber
 \theta = (-(T - T_C)/(T_{r1} - T_C)), T < T_C \\
 ((T - T_C)/(T_{r2} - T_C)), T > T_C,
 \end{eqnarray}
where $T_{r1}$ and $T_{r2}$ are the temperatures corresponding to $\Delta S^{max}_m$/$\Delta S_m$~=~0.5 below and above $T_C$, respectively for each applied field. Fig.~\ref{scaling} shows the Universal scaling plot for $\mathrm{Er_2Rh_3Si}$. It is noted that the $\Delta S_m$/$\Delta S^{max}_m$~$vs.$ $\theta$ curves nearly collapse onto a single master curve curve independent of temperature and field, indicating a second order magnetic phase transition in $\mathrm{Er_2Rh_3Si}$.

\section{SUMMARY}

The magnetic properties and magnetocaloric effect in the $\mathrm{Er_2Rh_3Si}$ compound have been studied in the present work. The ternary intermetallic compound $\mathrm{Er_2Rh_3Si}$ crystallizes in the rhombohedral $\mathrm{Mg_2Ni_3Si}$$\textendash$type structure and shows a paramagnetic to ferromagnetic phase transition with $T_C$ of 18 K. The explored MCE from magnetic isotherms revealed that $- \Delta S_m$, and RCP are 10.1 $\mathrm{J.~kg^{-1}.~K^{-1}}$, and 255 J. kg$^{-1}$ respectively for a change of magnetic field 0$\textendash$9 T. The existence of second order ferro$\textendash$paramagnetic phase transition is confirmed from Arrott plots and the universal scaling plots by normalized $\Delta S_m$ and rescaled temperature axis.

\section*{ACKNOWLEDGEMENT}   
This work is supported by Global Excellence and Stature (UJ-GES) fellowship, University of Johannesburg, South Africa. AMS thanks the URC/FRC of UJ for financial assistance.

\end{document}